\documentclass[apj]{emulateapj}
\usepackage{graphicx,url}
\usepackage{color}
\usepackage[backref,breaklinks,colorlinks,citecolor=blue]{hyperref}

\def\pcacorr{{\sc pcacorr}}

\shorttitle{Improving the Quality of {\it RXTE} HEXTE Spectra}
\shortauthors{Garc\'{\i}a et al.}

\begin{document}

\title{An Empirical Method for Improving the Quality of {\it RXTE} HEXTE Spectra}

\author{Javier~A.~Garc\'ia\altaffilmark{1}, 
        Victoria~Grinberg\altaffilmark{2},
        James~F.~Steiner\altaffilmark{1, 2},
        Jeffrey~E.~McClintock\altaffilmark{1}, 
        Katja~Pottschmidt\altaffilmark{3,4},
        Richard~E.~Rothschild\altaffilmark{5}}

\altaffiltext{1}{Harvard-Smithsonian Center for Astrophysics,
  60 Garden St., Cambridge, MA 02138 USA; javier@head.cfa.harvard.edu,
  jem@cfa.harvard.edu, jsteiner@head.cfa.harvard.edu}

\altaffiltext{2}{MIT Kavli Institute for Astrophysics and Space Research,
  MIT, 70 Vassar Street, Cambridge, MA 02139, USA; 
  grinberg@space.mit.edu}

\altaffiltext{3}{Department of Physics \& Center for Space Science and Technology, UMBC,
     Baltimore, MD 21250, USA}

\altaffiltext{4}{CRESST \& NASA Goddard Space Flight Center, Greenbelt, MD 20771, USA;
katja@milkyway.gsfc.nasa.gov}

\altaffiltext{5}{Center for Astrophysics and Space Sciences, University of California at San Diego, La Jolla, CA, USA,
rrothschild@ucsd.edu}

%

\begin{abstract}

  We have developed a correction tool to improve the quality of {\it
    RXTE} HEXTE spectra by employing the same method we used earlier to
  improve the quality of {\it RXTE} PCA spectra. We fit all of the
  hundreds of HEXTE spectra of the Crab individually to a simple
  power-law model, some 37~million counts in total for Cluster~A and
  39~million counts for Cluster~B, and we create for each cluster a
  combined spectrum of residuals. We find that the residual spectrum of
  Cluster~A is free of instrumental artifacts while that of Cluster\,B
  contains significant features with amplitudes $\sim1$\%; the most
  prominent is in the energy range 30--50~keV, which coincides with the
  iodine K edge. Starting with the residual spectrum for Cluster~B, via
  an iterative procedure we created the calibration tool {\sc hexBcorr}
  for correcting any Cluster~B spectrum of interest. We demonstrate the
  efficacy of the tool by applying it to Cluster~B spectra of two bright
  black holes, which contain several million counts apiece. For these
  spectra, application of the tool significantly improves the goodness
  of fit, while affecting only slightly the broadband fit
  parameters. The tool may be important for the study of spectral
  features, such as cyclotron lines, a topic that is beyond the scope of
  this paper.

\end{abstract}


%
%
%
%
\section{Introduction}\label{sec:intro}

The {\it Rossi X-ray Timing Explorer} ({\it RXTE}) was launched into a
low earth orbit on 30 December 1995 and operated continuously until the
mission was terminated on 4 January 2012. The three instruments aboard
{\it RXTE} were (i) the All Sky Monitor \citep[ASM;][]{lev96}, which
consisted of three coded aperture cameras that scanned about $\sim$80\%
of the sky every orbit; (ii) the Proportional Counter Array
\citep[PCA;][]{jah06}, a set of five proportional counter detectors
sensitive over the energy range 2--60~keV; and (iii) the High Energy
X-ray Timing Experiment \citep[HEXTE;][]{rot98}, which consisted of two
independent clusters (A and B), each with four NaI(Tl)/CsI(Na) phoswich
scintillation detectors sensitive over the energy range 15--250~keV. It
is the calibration of the latter instrument that is the focus of this
paper. A detailed discussion of the HEXTE detectors can be found in
\cite{rot98} and references therein; here, we provide a brief overview.

Each of the eight HEXTE detectors was fitted with a lead honeycomb
collimator giving a 1$^\circ$ FWHM field of view. All eight collimators
were co-aligned on source. The net open area of the eight detectors was
$\sim$1600\,cm$^2$ with an average energy resolution of 15.4\% FWHM at
60~keV. Both clusters A and B achieved a near-real-time estimate of the
background by being rocked between the source and a background field
through an angle of $1\fdg{}5$; the rocking axes of the clusters were
orthogonal. The exposure time on source was 32\,s, except early in the
mission when it was 16\,s. The corresponding observation times on the
background were 28\,s and 12\,s, respectively.

We improve the calibration of the HEXTE using precisely the same
approach we used previously for the PCA \citep{gar14b}. Namely, for each
cluster separately we fit individually all of the HEXTE spectra of the
Crab, which we assume to be featureless, to a simple power-law model. We
then combine the residual spectra to create two master spectra that have
extreme statistical precision. We find that the spectrum of Cluster~B
contains prominent instrumental artifacts, while the spectrum of
Cluster~A is essentially free of such artifacts. Via an iterative
process, we create the calibration tool {\sc hexBcorr}, and we
demonstrate the effectiveness of the tool in correcting the spectra of
two bright black holes. It is especially important to perform this
correction in studying spectra with high signal-to-noise, e.g., spectra
of bright sources and/or spectra created by combining several data
sets. We suggest that it may also be important for the study of spectral
features, such as cyclotron lines.

%
%
\section{Fits to Crab Spectra and the Creation of a Ratio
  Spectrum for Cluster~B}\label{sec:fits}

%
%
\begin{figure*}
\centering
\includegraphics[scale=0.6,angle=0]{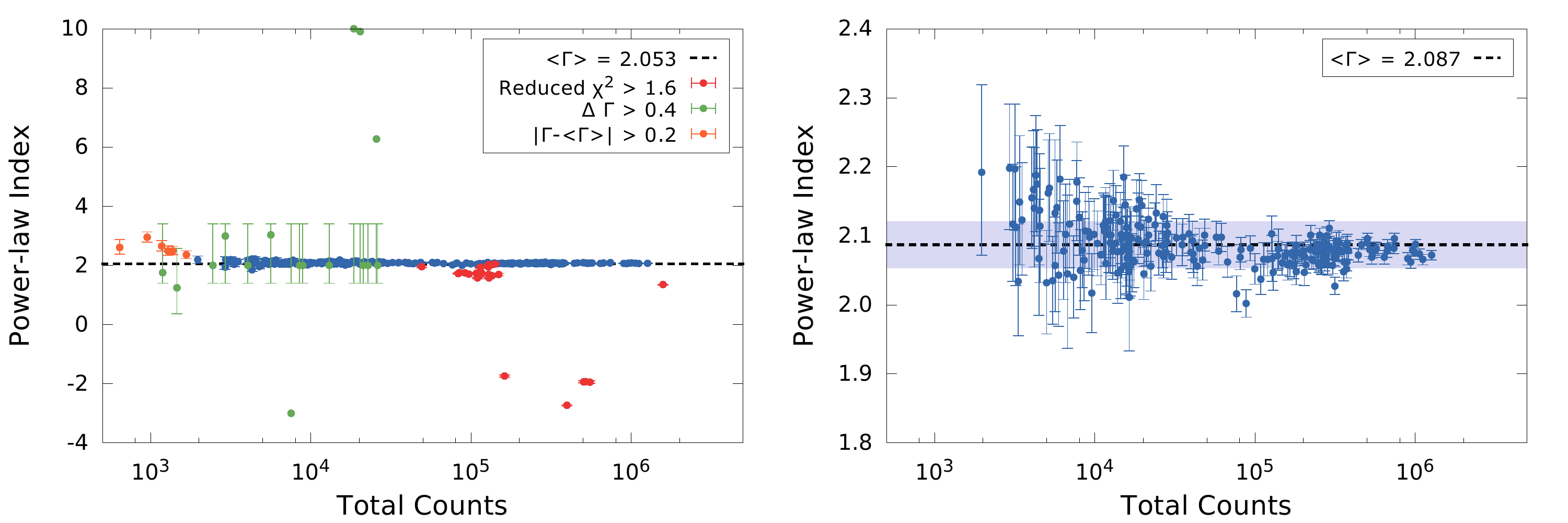}
\caption{
Photon power-law index vs.\ the number of counts in a Cluster~B
spectrum of the Crab. Results for the complete sample of 283 spectra are
shown in the left panel and results for the selected sample of 230 spectra
in the right panel. The dashed lines and shaded regions indicate the
average value and $\pm1$ standard deviation, respectively.
}
\label{fig:slope}
\end{figure*}
%

The performance of the HEXTE was affected by three major events during
the 16-year
mission\footnote{\url{http://heasarc.gsfc.nasa.gov/docs/xte/whatsnew/big.html}}.
(1) On 1996 March 6, the pulse height analyzer in one of the detectors
of Cluster~B failed so that after that date only three of the four
detectors were serviceable. (2) In October 2006 the mechanism that
rocked Cluster~A failed and the cluster was parked in the on-source
position. (3) In March 2010 the mechanism that rocked Cluster~B also
failed (having completed several millions cycles and far exceeded its
design goal). We consider only data that were taken when the instruments
were actively rocking because the analysis of HEXTE data that lack
quasi-simultaneous measurements of the background is problematic
\citep{pot06}.

Our analysis and discussion are focused on Cluster~B because, unlike
Cluster~A, it shows pronounced residual features; furthermore, it was
active $\approx4.3$~years longer than Cluster~A. For Cluster~A we
provide only a summary of results (Section~\ref{sec:ca}).

During the mission, 283 individual pointed observations of the Crab were
performed with HEXTE Cluster~B. All the spectra for both clusters have
been extracted using the standard tools in HEASOFT~6.16 and corrected
for deadtime using the {\tt hxtdead} tool. Visual inspection of the
data, preliminary power-law fits to the spectra, and information available
at the HEASARC revealed that for some observations the source was
occulted by the Earth or that the data were acquired in a non-standard
mode (e.g., the data corresponding to proposal P50100 lacks coverage
below $\sim$30~keV). Such data were excluded.

%
%
\begin{figure*}
\centering
\includegraphics[scale=0.8,angle=0]{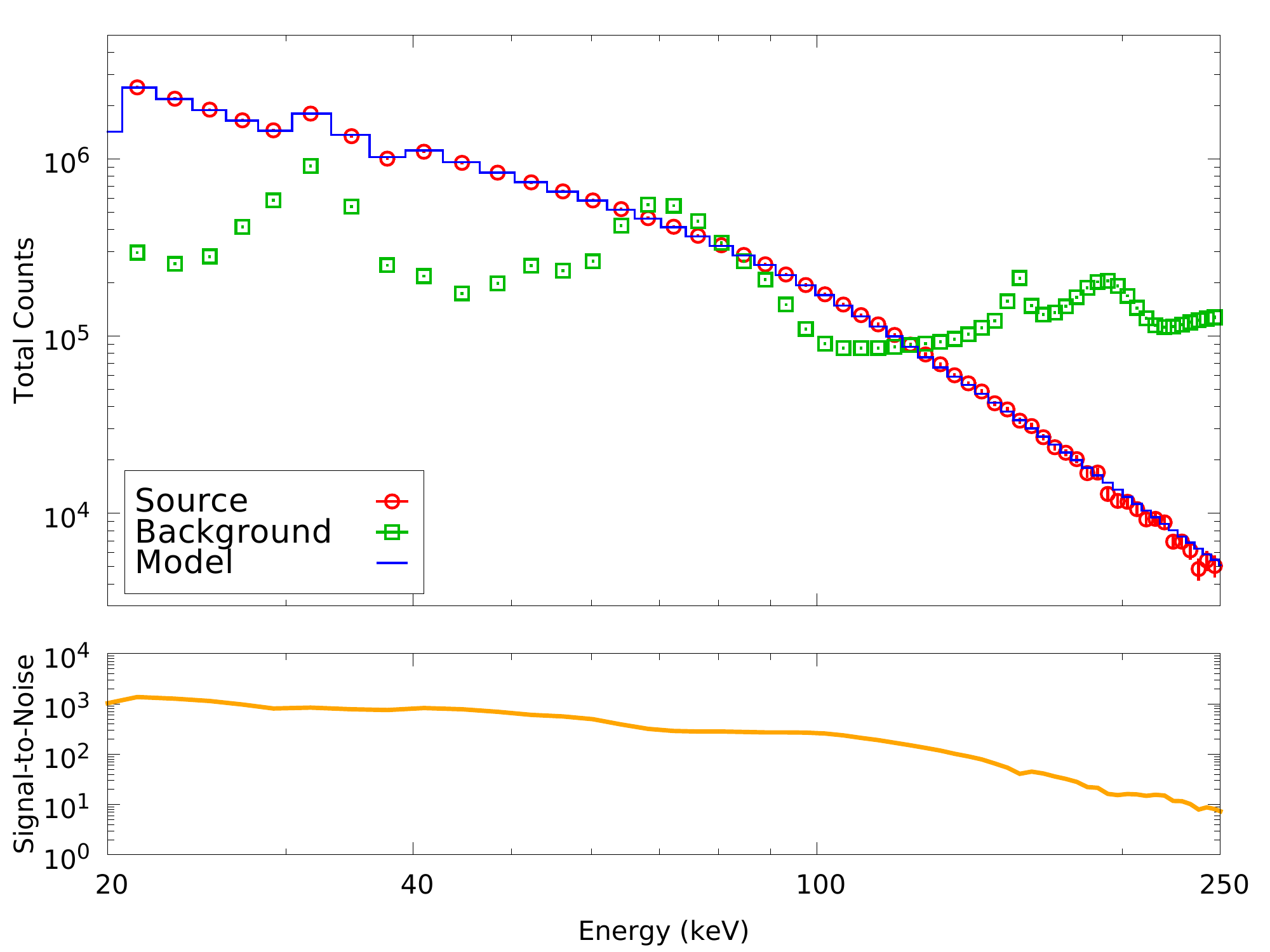}
\caption{ 
({\it top}) Total counts spectra for the source, background
  and model produced using all 230 Crab observations for HEXTE
  Cluster~B. ({\it bottom}) Significance of the signal, which for each
  channel individually is the total source counts divided by its
  uncertainty.  
}
\label{fig:data}
\end{figure*}
%

%
%
\begin{figure}
\centering
\includegraphics[scale=0.6,angle=0,trim={0.3cm 0 0 0}]{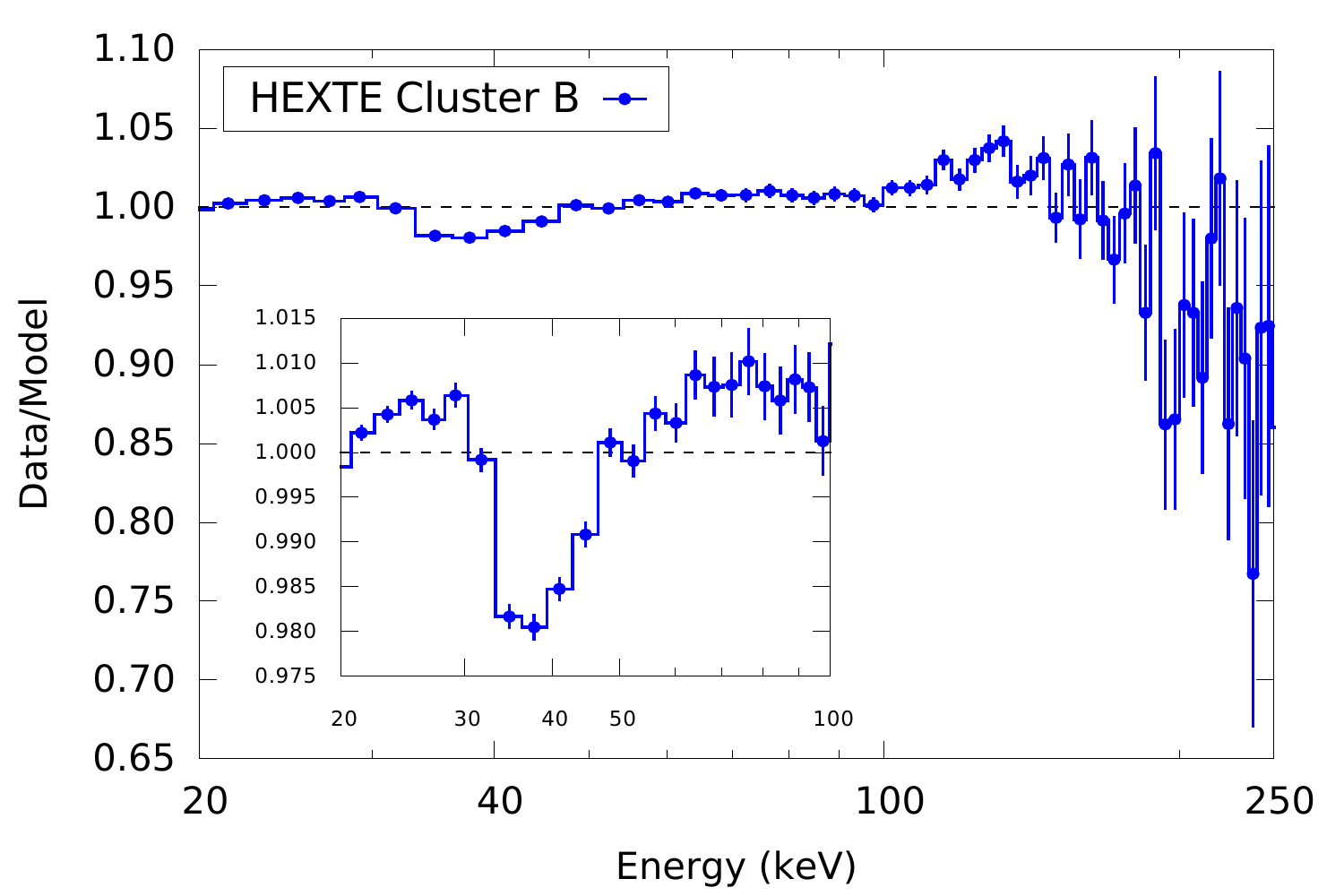}
\caption{
Ratio spectrum for Cluster~B created by combining the residual spectra
produced by fitting individually our complete sample of 230 selected
Crab spectra.
}
\label{fig:ratioCB}
\end{figure}

We analyzed each of the 283 observations separately. For all of our
model fitting and statistical analysis, we used {\sc xspec}~12.9.0d
\citep{arn96}. Working with the data as grouped by the standard
reduction procedure, we further binned the data to $\sim 3$ channels per
resolution element. Specifically, using {\sc ISIS}~1.6.2 \citep{hou02}
we binned up the data\footnote{Whether one bins the data using {\sc
    isis} or {\sc grppha} is unimportant; both tasks define the
  groupings in the PHA source file, while the background and response
  files remain unchanged.} by factors of 2, 3, and 4, in the energy
ranges 20--30~keV, 30--40~keV, and 40--250~keV, respectively.  No
allowance was made for systematic error in the response of the
detector. Each spectrum was fitted using a power-law model, with its
photon index $\Gamma$ and normalization as the only two fit
parameters. 

Because a break in the Crab spectrum has been reported by several
observers, we also alternatively fitted our data using a broken
power-law model. However, for nearly all of the spectra the break energy
was unconstrained. Furthermore, the combined ratio spectrum (discussed
below) differed only very slightly from that derived using the
single-slope model. For simplicity, we therefore adopted the unbroken
power-law model. For details, see Section~\ref{sec:disc}.

%
%
\begin{figure*}[!ht]
\centering
\includegraphics[scale=0.6,angle=0]{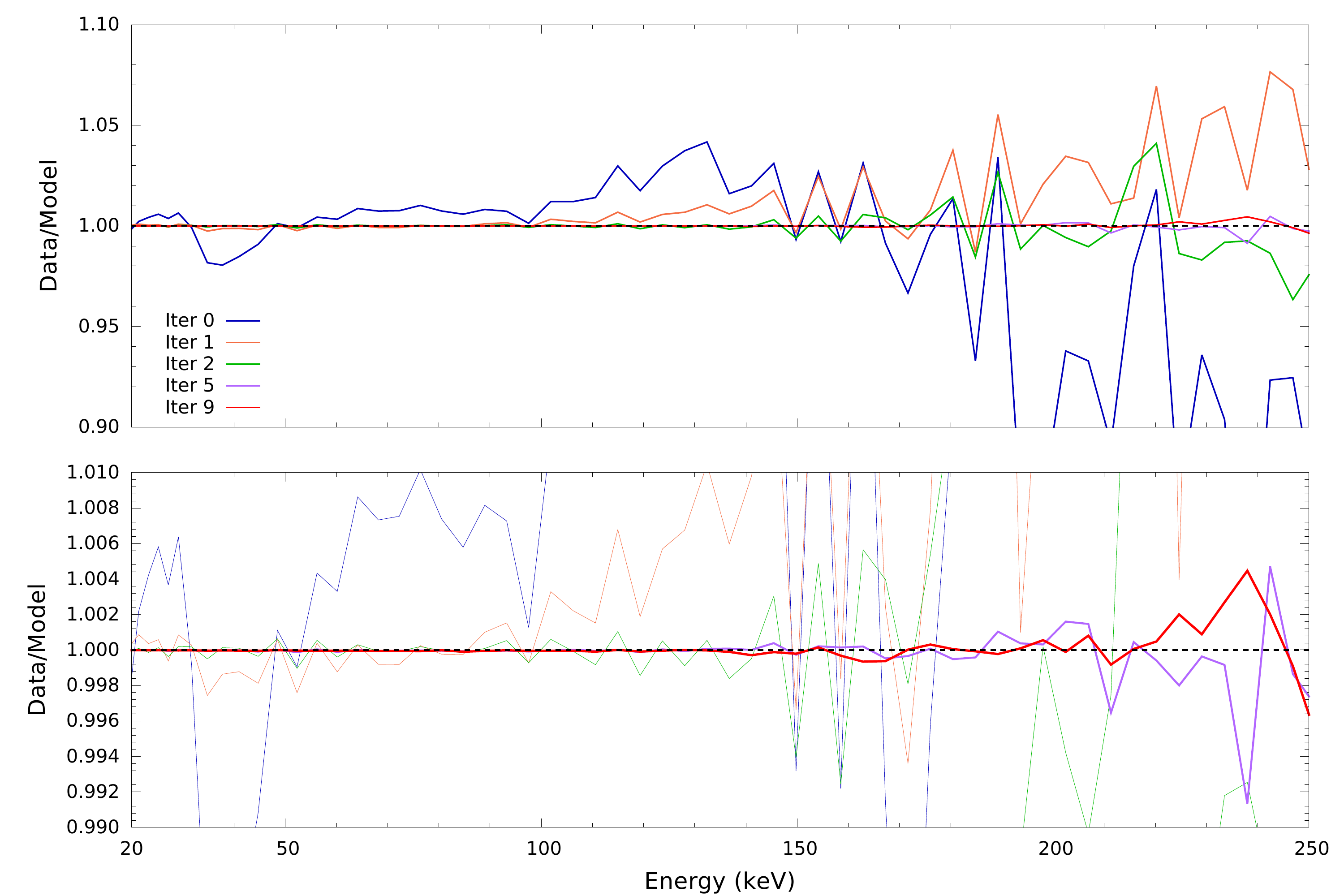}
\caption{
A succession of Cluster~B ratio spectra (which were created by the
iterative process described in the text) displayed over the full
20--250~keV band. ({\it top}) The parent ratio spectrum, the blue curve
labeled {\tt iter=0}, is identical to the ratio spectrum plotted in
Figure~\ref{fig:ratioCB}. Four of the nine smaller-amplitude daughter
spectra are shown, including that generated by the tenth and final
iteration (the red curve labeled {\tt iter=9}). ({\it bottom}) The same
ratio spectra more sensitively displayed.
}
\label{fig:iratio}
\end{figure*}

The left panel in Figure~\ref{fig:slope} shows for each of the 283
spectra the recovered power-law photon index $\Gamma$ vs.\ the total
number of counts. We discard three categories of data for which: the fit
is relatively poor with $\chi^2_\nu >1.6$ (red points); the uncertainty in
$\Gamma$ exceeds 0.4 (green points); and $\Gamma$ is more than 0.2 above
or below its mean value of $<\Gamma>=2.053$ (orange points). The poor
fits in the first category evidently result from background exposure
times that are anomalously short, typically only 3\% of the respective
source exposure time.

With the exclusion of these data, our final sample for Cluster~B is
comprised of 230 spectra which contains a total of 39~million
counts. The right panel in Figure~\ref{fig:slope} shows for these
selected spectra the power-law index vs.\ the total number of counts.
The modest variability in the photon index is comparable to that found
for the PCA by us \citep{gar14b} and by \cite{sha12}. We note that
long-term variations in the flux and photon index of the Crab have been
reported by \cite{wil11}. Attempting to corroborate these results is
beyond the scope of this paper. Meanwhile, these variations do not
affect our results (i.e., the performance of our correction tool)
because we fit each observation independently.

We now combine all 230 Cluster-B Crab spectra to produce the three ``total
counts spectra'' shown in Figure~\ref{fig:data}. The counts in
channel $i$ for the source spectrum is the sum over the individual
spectra $j$ of the background-subtracted source counts $S_i = \sum_j
S_{i,j}$.  The background and model spectra are similar sums over the
background counts ($B_i = \sum_j B_{i,j}$) and model counts ($M_i =
\sum_j M_{i,j}$). We emphasize that the model here is not a fit to the
summed spectrum, but rather it is the sum of the models fitted to the 230
individual spectra.  At energies $\gtrsim 140$~keV the background is
dominant. In the highest channel at 250~keV, there are $\sim100,000$
background counts and only $\sim5,000$ source counts.

The error bars for the source and background spectra in
Figure~\ref{fig:data} are plotted, but they are minuscule and scarcely
visible. For the source spectrum, the statistical uncertainty in the
number of counts in channel $i$ is
\begin{equation}\label{eq:sigma}
\sigma_i = \sqrt{ S_i + (T_o/T_b) B_i + (T_o/T_b)^2 B_i},
\end{equation}
where $T_o$ and $T_b$ are respectively the exposure times for the
observation of the source (and its background) and for the observation
of the background alone. The significance of the detection in each
channel (i.e., the signal-to-noise ratio), which is shown in the lower
panel of Figure~\ref{fig:data}, is simply
\begin{equation}\label{eq:sn}
(SNR)_i = S_i / \sigma_i.
\end{equation}

The ratio spectrum $S_i/M_i \pm \sigma_i/M_i$ is shown in
Figure~\ref{fig:ratioCB}. Its most distinctive feature is a $\approx1$\%
dip that extends from about 30~keV to 50~keV, which coincides with the
iodine K-edge at 33.17~keV \citep{way98}. Additional features that are
less significant are present above 100\,keV. At this point we could adopt
the ratio spectrum in Figure~\ref{fig:ratioCB} as a final product to be
used in correcting HEXTE spectra for Cluster~B. However, via the
iterative process described in the following section we obtain a final
product of much higher quality, which we refer to hereafter as the
``correction curve.''

%
%
\section{Correction Curve for Cluster~B and the Calibration Tool HEXBCORR}\label{sec:corr}

We now produce the correction curve for Cluster~B following precisely
the procedures we used earlier for the PCA, which are extensively
described in Section~4 in \cite{gar14b}. Producing a correction curve
for the HEXTE is simpler than for the PCA because the HEXTE automatic
gain control held the gain fixed throughout the mission so that every
observation has the same energy-to-channel mapping.

In brief: We start with the ratio spectrum shown in
Figure~\ref{fig:ratioCB}, which is identical to the curve labeled {\tt
  iter=0} in Figure~\ref{fig:iratio} except that the error bars have
been suppressed.  We then correct our 230 spectra by dividing each one
by this ratio spectrum and repeat the process described in the previous
section, thereby creating a new ratio spectrum that is labeled {\tt
  iter=1} in Figure~\ref{fig:iratio}. The procedure is repeated a total
of 10 times, resulting in the red curve labeled {\tt iter=9}, after
which additional iterations do not produce significant changes in the
curve. The panels in Figure~\ref{fig:iratio} show at two different
scales the reduction in the amplitude of the residual features achieved
at several points in the iteration process over the full energy range
(20--250~keV). The top panel highlights the gross improvements achieved
in the first few iterations, and the lower panel shows in detail how
each successive iteration reduces further the amplitude of every
residual feature in the spectrum.

We repeated the analysis described above, this time including a
correction for the normalization using the model {\tt recorn} in
{\sc xspec}. We find that the addition of this model component has a very
small effect on the ratio spectrum in Figure~\ref{fig:ratioCB}, producing
only mild effects ($\lesssim2\sigma$) at energies above 100~keV. The
effect is negligible in practice for the correction of HEXTE data, and
for simplicity we do not use {\tt recorn} in our analysis.

\begin{figure}
\centering
\includegraphics[scale=0.6,angle=0,trim={0.3cm 0 0 0}]{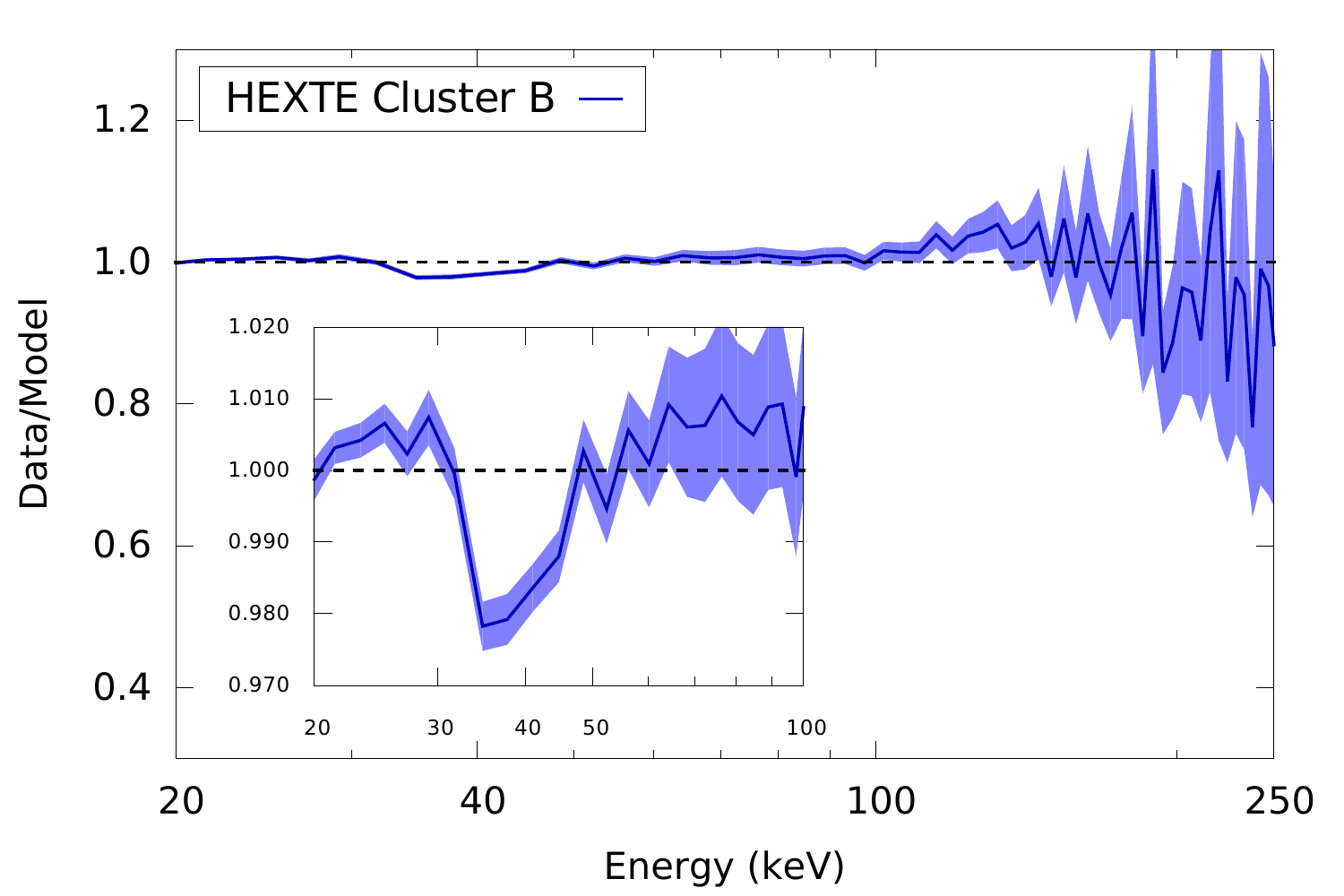}
\caption{
Final correction curve for Cluster~B. The lighter shaded region
bounding the curve shows the $1\sigma$ level of statistical error.
}
\label{fig:ratiohr}
\end{figure}

The final correction curve for Cluster~B is the product of all 10
correction curves (five of which are plotted in
Figure~\ref{fig:iratio}). This final correction curve, plotted in
Figure~\ref{fig:ratiohr}, constitutes the calibration tool {\sc
  hexBcorr}. To correct any Cluster~B object spectrum of interest
one simply divides the counts in each energy channel, as well as the
error, by the corresponding value of the correction curve.

%
%

%
%
\begin{figure*}
\centering
\includegraphics[scale=0.6,angle=0,trim={0.3cm 0 0 0}]{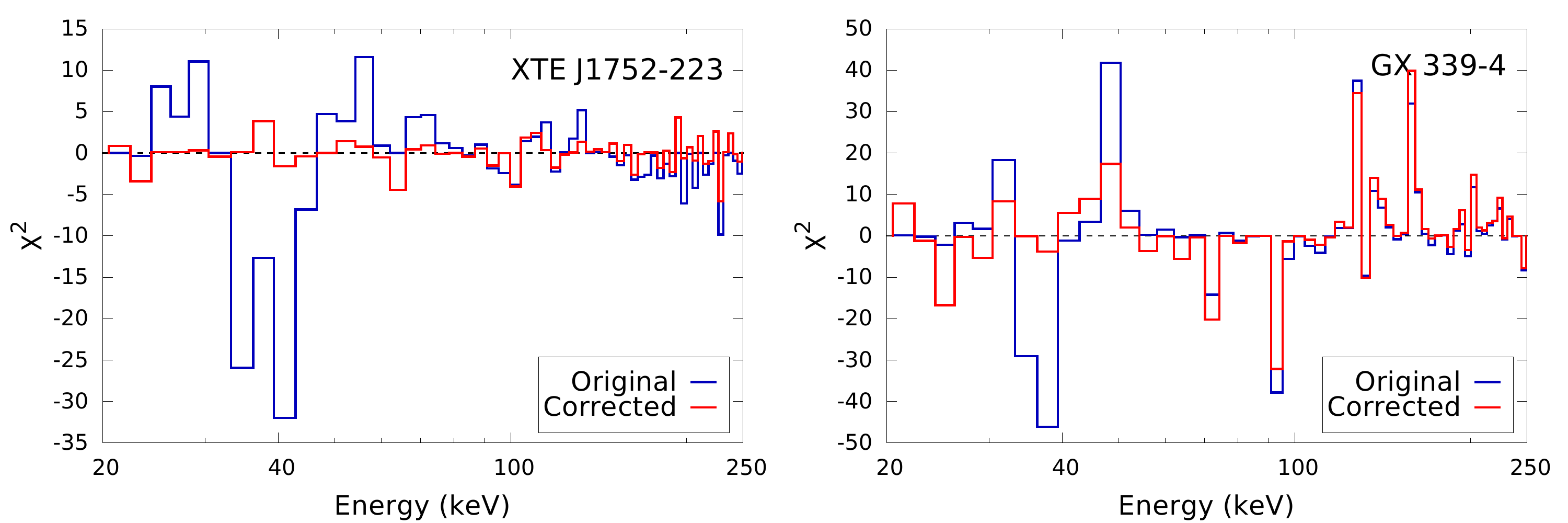}
\caption{
Comparison of the residuals from the fits to the original and the
corrected data for XTE~J1752--223 (left panel) and GX~339--4 (right
panel).
}
\label{fig:tests}
\end{figure*}

%
\section{Testing the Tool HEXBCORR on Spectra of Stellar-Mass
  Black Holes}\label{sec:test}

As a test of the calibration tool {\sc hexBcorr}, we apply it to
composite spectra of two bright transient Galactic black holes:
XTE~J1752$-$223 and GX~339$-$4. For both sources, the Cluster~B data
were collected in the bright hard state, and the spectra each contain
several million counts. Our analysis is aimed at demonstrating the
efficacy of {\sc hexBcorr}, and we do not concern ourselves with
employing an accurate physical model.

XTE~J1752$-$233: We selected 57 Cluster~B observations (all the data for
proposal numbers 94044 and 94331). We combined the spectra in two
steps. First, we created a summed residual spectrum precisely as we did
for the Crab, as described in Section~\ref{sec:fits}. Secondly, we then
added back in to this spectrum the average continuum component, which we
generated synthetically using the average values of the spectral index
and normalization parameter, and using the appropriate exposure time for
the summed spectrum. A detailed description of this process, and the
rationale for combining spectra in this way, is given in Section~3 of
\cite{gar15}. The summed spectrum contains a total of $10.4$\,million
counts in the energy range 20--250~keV. We applied the tool {\sc
  hexBcorr} to the spectrum by simply dividing the source counts, and
errors, by the correction curve shown in Figure~\ref{fig:ratiohr}. We
then fitted both the original (uncorrected) and the corrected spectra
using a cutoff power-law model. The residuals are compared in the left
panel of Figure~\ref{fig:tests}, which shows the contributions to
$\chi^2$ for each channel. The effect of the correction is clear-cut.
In particular, the strong 30--50~keV feature, present in the original
spectrum and in the Crab spectrum (Figure~\ref{fig:ratioCB}), is
completely absent in the corrected spectrum. Furthermore, the tool
reduces the residuals at almost every energy. The reduction in total
$\chi^2$ is striking: $\chi^2=205.1$ to $\chi^2=68.3$
($\Delta\chi^2=136.8$ with 54 degrees of freedom in both cases). At the
same time, the model parameters change only slightly: the photon index
from $1.23\pm0.01$ to $1.18\pm0.01$; the cutoff energy from
$125.7\pm2.4$~keV to $111.9\pm1.9$~keV; and the normalization from
$0.26\pm0.01$ to $0.23\pm0.01$ (where the first quantity is the value
for the original spectrum).

GX~339$-$4: We used the exceptionally bright hard-state Cluster~B data
collected in 21 observations made during 2002 April 20--30, which
correspond to the data defined by Box~A in \citet[][see their
Figure~1]{gar15}. The combined spectrum contains a total of $5.6$\,million
counts. We again fitted both the original and corrected data using the
same power-law model with a high energy cutoff; an additional mild
cutoff was required at low energies in order to achieve a good fit.  As
before, we do not seek a physical description of these data; rather we
apply a simple phenomenological model of the continuum.  The residuals
are compared in the right panel of Figure~\ref{fig:tests}, which shows
the contributions to $\chi^2$ for each channel. Once again, strong
residual features near 40~keV that are present in the uncorrected
spectrum are largely eliminated by the application of {\sc hexBcorr}.
The improvement in the fit is quite significant, although less so than
in the previous example: $\chi^2=392.1$ for the uncorrected spectrum and
$\chi^2=337.0$ for the corrected spectrum (i.e., $\Delta\chi^2 = 55.1$
with 53 degree of freedom in both cases), while the model parameters in
this case are consistent within the uncertainties. For the original and
corrected spectra, respectively, the photon index is $1.65\pm0.03$ and
$1.60\pm0.03$; the high energy cutoff is $73.0\pm2.5$ and
$67.0\pm2.2$; and the normalization is $2.3\pm0.2$ and
$2.1\pm0.2$. In comparison with the results for XTE~J1752$-$223, the
consistency of the fit parameters and smaller value of $\Delta\chi^2$
for GX~339$-$4 can be reasonably explained by noting that its spectrum
contains only about half as many counts.

%
\section{Fits to Crab Spectra and the Creation of a Ratio Spectrum for Cluster~A}\label{sec:ca}

\begin{figure}
\centering
\includegraphics[scale=0.6,angle=0,trim={0.3cm 0 0 0}]{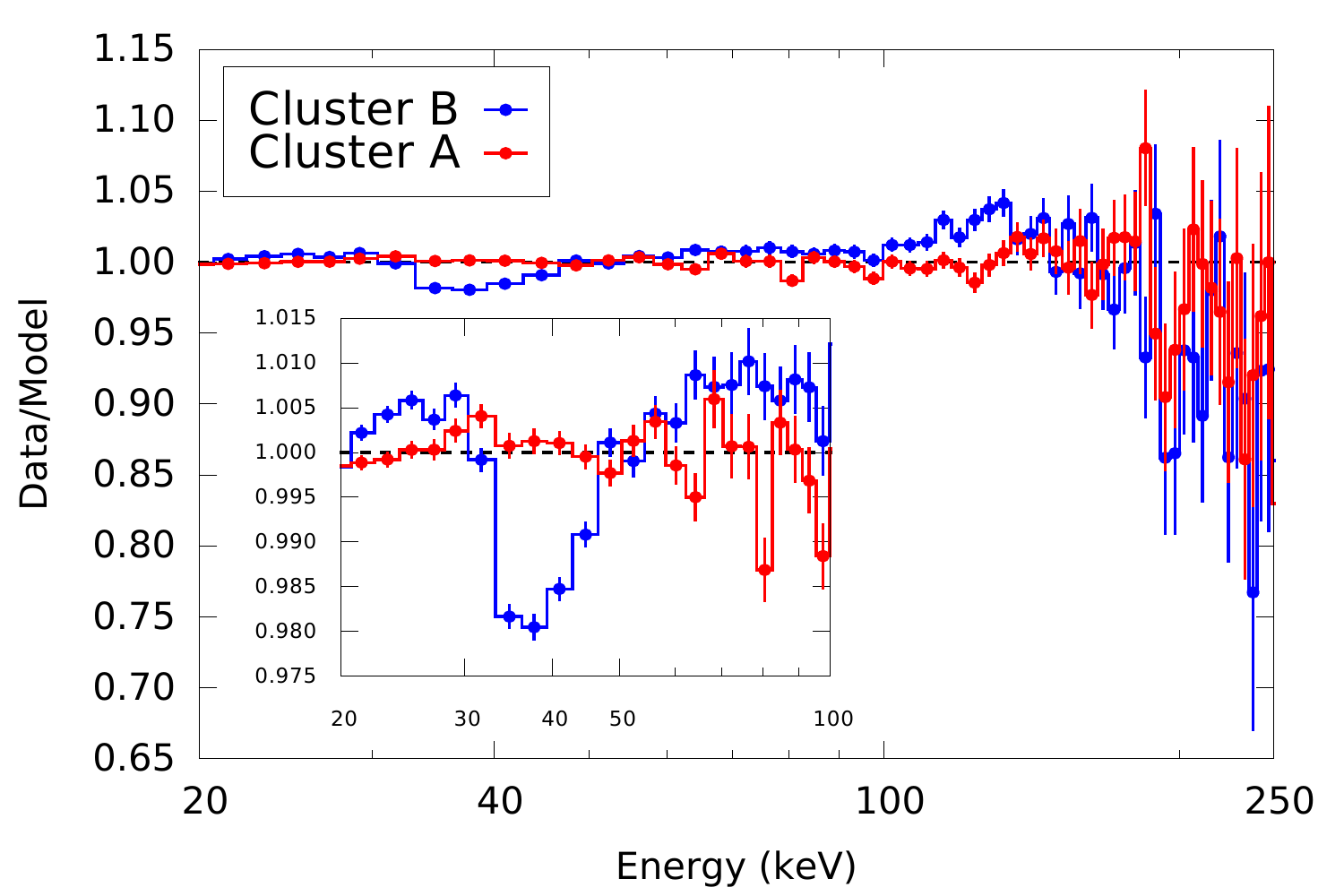}
\caption{
Comparison of the residuals for the two clusters from fits to the
Crab data. The blue spectrum for Cluster~B is identical to that shown in
Figure~\ref{fig:ratioCB}.
}
\label{fig:ratioCAB}
\end{figure}

We performed precisely the same global analysis of the Crab data for
Cluster~A that we performed for Cluster~B. As in Section~\ref{sec:fits},
we fitted all the available data for Cluster~A, in this case 204
observations; fitted the spectra with a simple power-law model; applied
the same selection criteria; and arrived at our final data sample of 168
spectra comprising a total of 37\,million counts. Again, following the
procedures described in Section~\ref{sec:fits}, we produced a
data-to-model ratio spectrum that is a sum of all the selected
data. This ratio spectrum for Cluster~A is compared to that of Cluster~B
in Figure~\ref{fig:ratioCAB}.  There are no significant residual
features in the spectrum of Cluster~A, which is an unexpected result
given that the clusters were built to the same design.  The residuals at
all energies are approximately consistent with counting statistics. We
conclude that it is unnecessary to correct Cluster~A data.

\begin{figure}
\centering
\includegraphics[scale=0.6,angle=0,trim={0.3cm 0 0 0}]{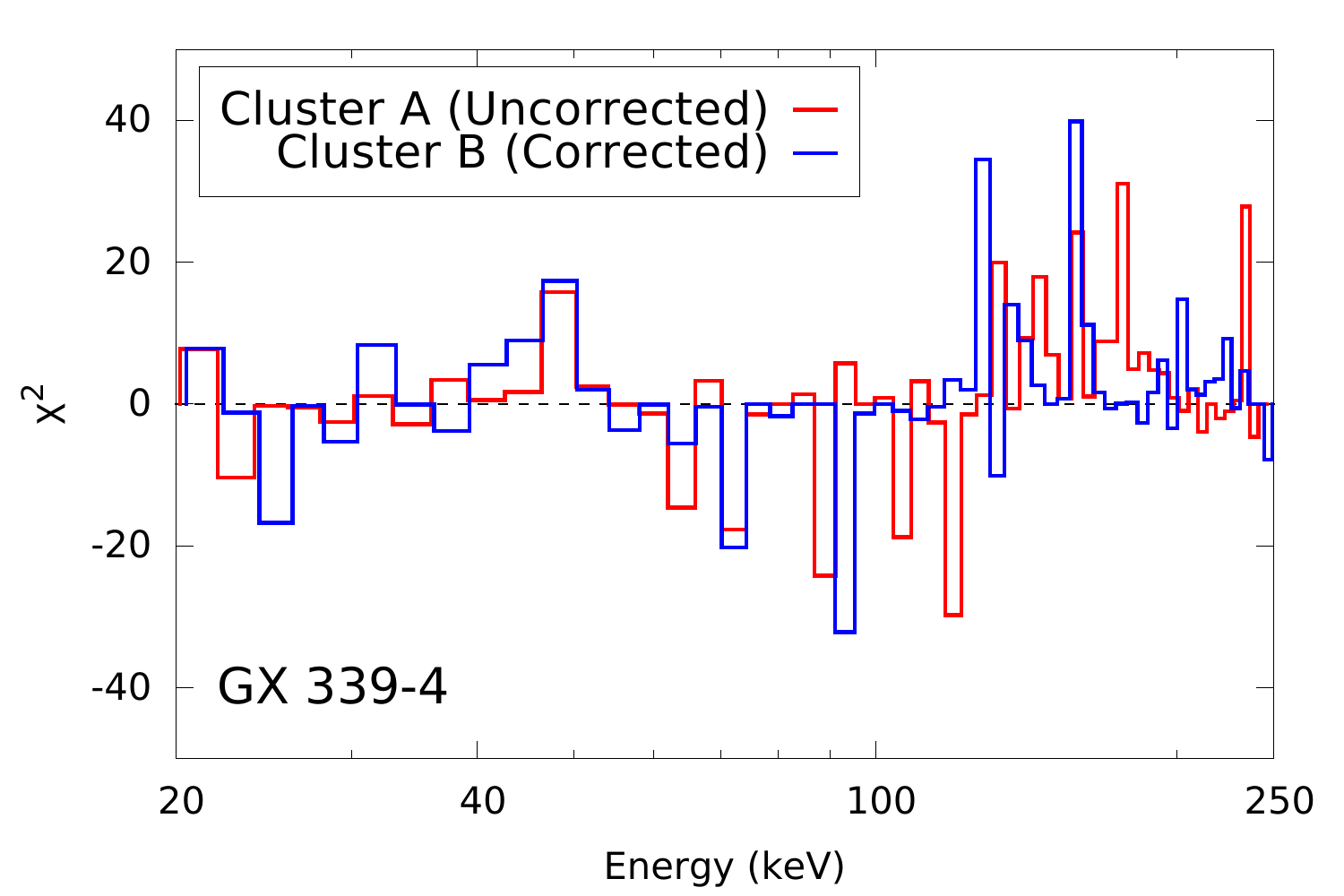}
\caption{
Comparison of the residuals for the two clusters from fits to the
combined spectrum of GX~339-4 with 5.6\,million counts.
}
\label{fig:gx339-CAB}
\end{figure}

We tested the quality of raw (i.e., uncorrected) Cluster~A data for
GX~339--4 by making a direct comparison with the corrected Cluster~B
spectrum of this source shown in Figure~\ref{fig:tests}. (Note that no
rocked Cluster~A data are available for XTE~1752$-$233; see Section~\ref
{sec:fits}.) We generated the Cluster~A spectrum of GX~339--4 using
precisely the same procedures used for the Cluster~B spectrum
(Section~\ref{sec:test}), and we fitted both spectra independently using
the same cutoff power-law model. The fit residuals for the two
spectra are compared in Figure~\ref{fig:gx339-CAB}. The quality of the
fits is very similar, and all the model parameters are consistent within
the uncertainties. The total $\chi^2$ is slightly worse for the
Cluster~A spectrum ($\chi^2=332.7$ vs.\ $\chi^2=371.43$), but this is
likely largely because the number of counts is greater (7.7\,million
vs. 5.6\,million). Thus, we find that the uncorrected Cluster~A data are
comparable in quality to the corrected Cluster~A data, confirming our
conclusion above that it is unnecessary to correct Cluster~A data.

%
%

%
%

\section{Discussion and Conclusions}\label{sec:disc}

Cluster~A data are free of the instrumental features that require
correction in the Cluster~B data.  This is surprising if these features
are instrumental because one would expect the performance of these
essentially identical clusters to be more nearly similar. While the
reason for the dissimilarity is unclear, the leading hypothesis is that
it resulted from adjusting Cluster~B's calibration two months into the
mission, which was necessitated by the failure of one of its four
detectors (see Section~\ref{sec:fits}). Unfortunately, the Crab data
collected prior to the failure are too sparse to corroborate this
hypothesis. Since the Brookhaven calibration was done with a single HEXTE
detector, it is possible that one or more cluster B detectors had a slightly
different response at and just beyond the K-edge of iodine, and that is the
source of the residual, since the same two segment description of the edge
response was used for all detectors.

The calibration of the HEXTE depends on both laboratory and in-flight
data. The efficiency of the detectors vs.\ energy was determined prior
to launch using radioactive sources and monochromatic X-rays generated
at Brookhaven National Laboratory. This part of the instrument response
includes the energy-dependent escape of photons above the K edge of
iodine. The efficiency vs.\ energy above and below the edge was mimicked
by two line segments, which is an imperfect model because the profile of
the edge is more complicated than a step function.

The final calibration of the open area and point spread function
(PSF) of the instruments was determined post-launch using Crab
data. Multiple observations were made on-axis and over a range of
off-axis angles. These data were used to adjust the preliminary
laboratory measurements of the PSF and open area. None of these
individual Crab observations was sensitive enough to detect the 1\% dip
seen in the combined ratio plot (Figure~\ref{fig:ratioCB}). The
adjustments to the calibration of Cluster~B after the detector failure
were solely to the open area and the PSF; the detector efficiencies were
left unchanged. These adjustments are the most significant event that
differentiates the two HEXTE clusters. However, it remains an open
question precisely how this event could produce the relatively large residuals in
Cluster~B that are absent in Cluster~A.

%
%

As noted in Section~\ref{sec:fits}, in creating the tool {\sc hexBcorr}
we analyzed all the Crab data using a simple absorbed power-law model
while ignoring the evidence that the spectrum breaks at an energy that
is quite uncertain: $79\pm10$~keV \citep{str79}, $128\pm4$~keV
\citep{jun89}, $60\pm7$~keV \citep{bar94} $57\pm3$~keV \citep{rot98},
$100$~keV (fixed) \citep{jou09}, $105\pm20$~keV and $117\pm19$~keV
\citep{yam11}.  In addition to these reports of a break, we note that
our ratio spectrum in Figure~\ref{fig:ratioCB} and our final correction
curve in Figure~\ref{fig:ratiohr} show marginal evidence for a break at
energies $\gtrsim150$~keV, although this apparent defect in our
correction curve is relatively unimportant because spectra contain few
source counts at these energies and the background is the dominant
source of uncertainty.

As an alternative to the simple power-law model, we fitted our 283
Cluster-B spectra using a broken power-law model, but we obtained
unsatisfactory results; namely, for nearly all the spectra the break
energy was unconstrained with values scattered randomly across the
allowed range of 50--200 keV. Most importantly, the net ratio plot
for these fits was only marginally different from that obtained using
the simple power-law model, which accurately captures the important
$\sim1$\% dip at $\sim40$~keV and the other principal features in
the ratio spectrum (Figure~\ref{fig:ratioCB}). Therefore, all of our
results are based on the simple power-law model.

Our results for XTE~J1752--223 and GX~339--4 indicate that for combined
spectra of bright sources, which are comprised of dozens of individual
spectra and contain several million counts, the application of {\sc
  hexBcorr} significantly improves the quality of the fit to Cluster~B
data, while having at most a modest effect on broadband spectral
parameters. For example, for the latter source with 5.6~million counts
the values of the photon index, cutoff energy, and normalization were
consistent within the statistical uncertainties. For XTE~J1752--223 with
10.4~million counts, the parameters changed modestly, but by several
standard deviations: the photon index, cutoff energy and normalization
changed by $\Delta\Gamma=0.05\pm 0.01$, $\Delta E=13.8\pm3.1$~keV, and
$\Delta N=0.03\pm0.01$.  For an individual spectrum of a bright source
with an exposure time of only a few ks, it is likely that the correction
is at most cosmetic with essentially no effect on the broadband
parameters.

However, the correction is potentially important for the study of
spectral features such as cyclotron lines, particularly those at
energies of $\sim 30-50$~keV. There have been a number of reports of
cyclotron lines in this band, some based on HEXTE data
\citep[e.g.,][]{hei99,cob01, hei01,hei03,rod09,tsy12}. The instrumental
features we detect have an amplitude of about 1\%, which is much less
than that of a typical cyclotron line. However, in some cases the
application of our calibration tool may prove fruitful.  For example,
\cite{dec13}, who discovered a 10~keV cyclotron line (with an amplitude
$\sim3\%-10$\%) in the spectrum of Swift~J1626.6--5156, find residual
features in their PCA and HEXTE (Cluster~B) spectra at $\sim 40
$~keV. The authors argue that these features are instrumental and not a
harmonic of the 10~keV line, an hypothesis that can possibly be tested
using our calibration tool. A spectrum of the bursting pulsar
GRO~J1744--28 provides a second example where a residual feature was
interpreted as an instrumental artifact rather than as a cyclotron line
\citep{hei99b}. The question of whether our calibration tool is actually
important for the study of cyclotron lines is beyond the scope of this
paper.

\begin{figure}
\centering
\includegraphics[scale=0.6,angle=0,trim={0.4cm 0 0 0}]{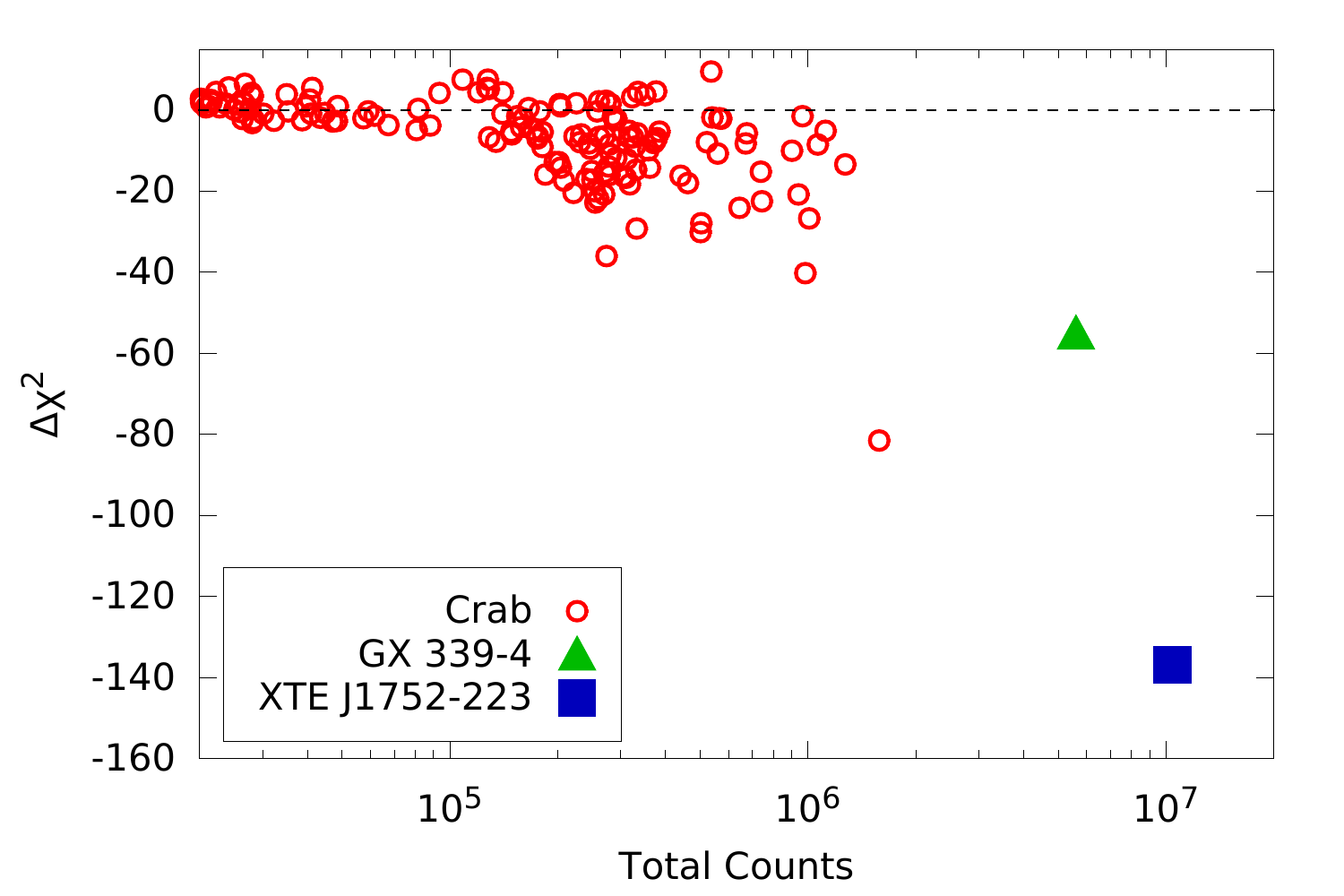}
\caption{ 
Improvement in the fit for Cluster-B data, as measured by a
  decrease in $\chi^2$, as a function of total counts for bright spectra
  of the Crab and for summed spectra of GX~339--4 and
  XTE~J1752--224. The Crab spectra were fitted over the 20--250~keV band
  using the simple absorbed power-law model. The results shown for the
  two black hole sources are those given in Section~\ref{sec:test}.  
}
\label{fig:deltachi}
\end{figure}

The benefits of correcting HEXTE Cluster-B data with {\sc hexBcorr} are
greatest for spectra with many counts. Figure~\ref{fig:deltachi} roughly
quantifies this benefit by plotting the improvement in the fit achieved
by application of the tool to the brighter Crab spectra as a function of
the total counts. The effect of the correction becomes apparent for
spectra with $\sim10^5$ counts and it becomes quite significant as the
number of counts approaches $10^6$. The improvement in the fit for the
spectra of GX~339--4 and XTE~J1752--228 with several million counts
apiece is dramatic. The figure indicates the value of making the
correction for spectra with more than $\sim10^5$~counts. This limit
should be considered only a useful rule of thumb since it likely depends
on the spectral shape.

%
%

In summary, we have demonstrated that a mission-averaged spectrum of the
Crab with 39~million counts in the HEXTE band reveals imperfections in
the calibration of Cluster~B. Following a procedure designed originally
for correcting PCA data, we show how to reduce the principal
$\sim$30--50~keV instrumental feature in Cluster~B data by an order of
magnitude, while significantly reducing the residuals at nearly all
energies in the full 20--250~keV band. To correct any Cluster~B spectrum
of interest one applies the tool {\sc hexBcorr}, which divides the
spectrum, channel-by-channel, with the correction spectrum shown in
Figure~\ref{fig:ratiohr}. We show that for combined spectra of bright
sources containing more than $10^5$ counts, the correction greatly
improves the quality of the fit while only mildly affecting the
broadband fit parameters. For individual spectra of bright sources with
many fewer counts, the effects on the broadband parameters will be
correspondingly less. However, for the study of discrete spectral
features at energies of $\sim$30--50~keV, such as cyclotron lines, the
correction may be important even for individual spectra. Finally, we
find no significant residual features in the combined Crab spectrum
using Cluster~A observations, and we conclude that no correction is required
for these data. Earlier, we made publicly available the calibration tool
\pcacorr\ for correcting PCA data \citep{gar14b}. Now, the correction
curve for HEXTE Cluster~B along with a Python script, which constitutes
the tool {\sc hexBcorr}, is publicly available at 
\url{http://hea-www.cfa.harvard.edu/~javier/hexBcorr/}.

%
%
%
\acknowledgments 
We thank an anonymous referee for several helpful comments.
JG and JEM acknowledge the support of a CGPS grant from
the Smithsonian Institution. JFS has been supported by NASA Hubble
Fellowship grant HST-HF-51315.01 and NASA Einstein Fellowship grant
PF5-160144. VG acknowledges support provided by NASA through the
Smithsonian Astrophysical Observatory (SAO) contract SV3-73016 to MIT
for support of the Chandra X-Ray Center (CXC) and Science Instruments;
CXC is operated by SAO for and on behalf of NASA under contract
NAS8-03060.
%
%
%
%
\bibliographystyle{apj}
\bibliography{my-references}
%
%
%
%
\end{document}